\documentclass[12pt]{article}
\usepackage{amsmath}
\usepackage{graphicx}

\begin{document}

\begin{center}

{\bf\large{Unstable Particles in Quantum Field Theory}}\\

\bigskip

Scott Willenbrock \\

\bigskip

Department of Physics, University of Illinois at Urbana-Champaign \\
1110 West Green Street, Urbana, Illinois 61801, USA \\

\bigskip

willen@illinois.edu

\end{center}

\medskip

\noindent {\bf Abstract}: In honor of Dave Roper's $90^{\rm th}$ birthday, I present a pedagogical introduction to our modern understanding of unstable particles in Quantum Field Theory, based on the analytic structure of the propagator, with occasional remarks on the Roper resonance. I discuss the mass and decay rate of unstable particles, Breit-Wigner resonance formulae and width, poles and branch cuts, and pole trajectories.

\bigskip\medskip

\noindent {\bf 1.\ Introduction}

\bigskip

I would like to take this opportunity, on the occasion of Dave Roper's $90^{\rm th}$ birthday, to review our modern understanding of unstable particles in Quantum Field Theory \cite{W1}. You might think that this could be found in any one of the many excellent textbooks on Quantum Field Theory, but that is not the case. All such textbooks either avoid the topic altogether, or treat it approximately (sometimes without acknowledging the approximations).\footnote{See Sec.~7 of Ref.~\cite{W2} for a review of the treatment of unstable particles in Quantum Field Theory textbooks.} The treatment here is close to that of Sec.~6.3 of the 1992 textbook of Lowell Brown \cite{Brown}, but without assuming the width of the particle is small compared to its mass.

The Roper resonance, along with the Higgs boson, is one of the only particles that carries someone's name. This baryon, discovered by Dave in 1964 in a careful analysis of pion-nucleon scattering data \cite{Roper}, was among the first hadronic resonances discovered. Its discovery was a surprise because it was not conspicuously apparent in the data. It is the lightest baryon resonance with the same quantum numbers as the nucleon ($J^P = \frac{1}{2}^{+}$, $I=\frac{1}{2}$), is the second lightest of the dozens of unflavored baryon resonances (after the $\Delta(1232)$), and is still of considerable interest today. It is labeled $N(1440)$ by the Particle Data Group \cite{RPP2024}, but researchers in the field still call it the Roper resonance.

My goal is to give a treatment that applies to all unstable particles, whether it be a muon, a pion, a $B$ meson, a hadronic resonance, or a $Z$ boson. Although these particles manifest themselves in experiments in very different ways, they are all treated the same way in Quantum Field Theory. A muon, charged pion, and $B$ meson decay weakly, have widths many orders of magnitude less than their masses, and travel macroscopic distances before decaying. A hadronic resonance decays strongly, so its width is typically $\cal{O}\rm{(100)\; MeV}$, just one order of magnitude less than its mass, corresponding to a lifetime of about $10^{-23}$ sec. A $Z$ boson decays weakly, but because it is so heavy its width is about 2.5 GeV, roughly two orders of magnitude less than its mass, corresponding to a lifetime of about $10^{-24}$ sec.

The treatment of unstable particles here is pedagogical, with a minimum of references, in the style of a textbook. Extensive references to the literature and further details can be found in Refs.~\cite{W1,W2}. I hope that the treatment of unstable particle that I present here will inform the next generation of Quantum Field Theory textbooks.

\bigskip

\noindent {\bf 2.\ Mass}

\bigskip

Let us begin by asking why we are using Quantum Field Theory to describe unstable particles. Perhaps this is obvious for the muon and the $Z$ boson, whose fields are present in the electroweak Lagrangian, but what about the pion, the $B$ meson, and the hadronic resonances? After all, these particles are composed of quarks and antiquarks, and the QCD Lagrangian is written in terms of quarks fields, not baryon and meson fields. The answer is that if we want to describe the interactions of hadrons in a way that is consistent with quantum mechanics and relativity, then we must use Quantum Field Theory. The hadron fields are certainly not fundamental, but they are nevertheless useful to describe hadron interactions. This is the point of view of effective field theory \cite{Weinberg}.

A stable particle is described by a one-particle state that is an eigenstate of the energy-momentum operator (see Sec.~2.5 of Ref.~\cite{Weinberg})
\begin{equation}
P^\mu \Psi_p = p^\mu \Psi_p
\end{equation}
where $p^2 = m^2$ is the squared mass of the particle. The zeroth component of the operator is the Hamiltonian, $P^0 = H$, whose eigenvalue is the energy. Applying the time-translation operator $e^{-iHt}$ to the state yields $\Psi^{\prime}_p = e^{-iEt}\Psi_p$, which is the familiar time dependence of any energy eigenstate.

An unstable particle does not correspond to an asymptotic state, so it cannot be described by a one-particle state.\footnote{If the particle's lifetime is much greater than the interaction time, then it can be approximated as an asymptotic state. We are interested in the general case with no approximations.} Instead, we must create the unstable particle and watch it propagate in time to learn about its mass and decay rate.

For simplicity, let's consider a spin-zero particle and its associated complex scalar field $\phi(x)$. Let's begin with a free field; we will add interactions momentarily. The field destroys particles and creates antiparticles at spacetime position $x$, while its conjugate does the opposite. The amplitude for a particle to be created at the origin and destroyed at $x$ (or an antiparticle to be created at $x$ and destroyed at the origin) is the propagator,
\begin{equation}
\langle 0 | T \phi(x) \phi^\dagger (0) |0\rangle = \int \frac{d^4p}{(2\pi)^4}\frac{ie^{-ip\cdot x}}{p^2-m^2+i\epsilon}
\end{equation}
where $m$ is the physical mass of the free particle.

In order for the particle to decay, we must add interactions. Fortunately it is not hard to include interactions to all orders in perturbation theory by summing the series of self-energy diagrams shown in Fig.~1, where the shaded circle denotes the bare one-particle-irreducible self energy, $i\Pi(p^2)$.
Summing the series of diagrams gives
\begin{eqnarray}
&&\frac{i}{p^2 - m_B^2+i\epsilon}\left(1-\frac{\Pi(p^2)}{p^2-m_B^2+i\epsilon}+\cdots\right)\nonumber\\
&&=\frac{i}{p^2 - m_B^2 + \Pi(p^2)+i\epsilon}
\label{bare}
\end{eqnarray}
where $m_B$ is the bare mass.
The pole of the full momentum-space propagator, $\mu^2$, is the solution to the equation
\begin{equation}
\mu^2-m_B^2+\Pi(\mu^2)=0\;.
\label{pole}
\end{equation}
Combining Eqs.~(\ref{bare}) and (\ref{pole}) yields
\begin{equation}
\langle 0 | T \phi(x) \phi^\dagger (0) |0\rangle = \int \frac{d^4p}{(2\pi)^4}\frac{ie^{-ip\cdot x}}{p^2-\mu^2+\Pi(p^2)-\Pi(\mu^2)+i\epsilon}
\label{bare2}
\end{equation}
where we have eliminated the bare mass in favor of $\mu$. If $\mu$ is real, this just amounts to ordinary mass renormalization of a stable particle, and $\mu$ is the physical mass. If $\mu$ is complex, however, the particle is unstable, and the interpretation of $\mu$ requires further analysis.\footnote{Since the unstable particle is not an asymptotic state it is not necessary to renormalize the field, although one may choose to do so.}

The propagator by itself is unphysical; one cannot simply create a particle from the vacuum. Rather, the propagator is an ingredient in the construction of a scattering amplitude. The scattering amplitude acquires a complex pole from the propagator of an intermediate unstable particle,\footnote{The propagator also introduces a branch cut; we will discuss this in Sec.~6.}
\begin{equation}
{\cal M} \sim \frac{1}{p^2-\mu^2}\;.
\label{LOprop}
\end{equation}
Going to the rest frame of the unstable particle we can rewrite Eq.~(\ref{LOprop}) as
\begin{equation}
{\cal M} \sim \frac{1}{p_0^2-\mu^2}=\frac{1}{(p_0 - \mu)(p_0 + \mu)}\;.
\label{Epole}
\end{equation}
To find the time dependence of the scattering amplitude, we (inverse) Fourier transform from energy to time, that is, we perform the $p_0$ integral of Eq.~(\ref{bare2}) while setting ${\bf p}=0$:
\begin{equation}
{\cal M} \sim \int_{-\infty}^\infty \frac{dp_0}{2\pi}\;  \frac{ie^{-ip_0t}}{(p_0 - \mu)(p_0 + \mu)}\;.
\label{FT}
\end{equation}
For $t>0$ we can close the contour in the lower half complex $p_0$ plane, as shown in Fig.~2, since the integral along the large semicircle is exponentially damped. Using the residue theorem, we pick up the contribution of the pole at $p_0 = \mu$,
\begin{equation}
{\cal M} \sim e^{-i\mu t} = e^{-i{\rm Re}\,\mu t}e^{{\rm Im}\,\mu t}\;.
\label{time}
\end{equation}
The probability is given by the square of the scattering amplitude,
\begin{equation}
|{\cal M}|^2 \sim e^{2{\rm Im}\,\mu t}\;.
\end{equation}
This corresponds to exponential decay with $2{\rm Im}\,\mu = -\Gamma$, where $\Gamma$ is the decay rate. Hence we learn that ${\rm Im}\,\mu = - \Gamma/2$.

For $t<0$ we close the contour in the upper half complex $p_0$ plane and pick up the contribution of the pole at $p_0 = -\mu$. This pole is associated with antiparticle propagation. The residue theorem gives ${\cal M} \sim e^{i{\rm Re}\,\mu t}e^{-{\rm Im}\,\mu t}$, and hence $|{\cal M}|^2 \sim e^{\Gamma t}$. This also corresponds to exponential decay since $t$ is negative, and proves that the particle and antiparticle have the same lifetime, a result that holds for any Quantum Field Theory.

The oscillatory frequency of the unstable particle is dictated by the particle energy, $e^{-iEt}$, as for a stable particle. In the particle's rest frame its energy is just its mass. Hence ${\rm Re}\,\mu = m$, and we conclude that
\begin{equation}
\boxed{\mu = m - \frac{i}{2}\Gamma}\;.
\label{mGamma}
\end{equation}
Eq.~(\ref{mGamma}) provides an unambiguous decomposition of $\mu$ into physically meaningful quantities. For this reason we will refer to $m$ as the physical mass.

As a sanity check, let's evaluate the Fourier transform of the scattering amplitude, in a frame with nonzero three-momentum ${\bf p}$:
\begin{equation}
 {\cal M} \sim \int_{-\infty}^\infty \frac{dp_0}{2\pi}\;  \frac{ie^{-ip_0t}}{\left(p_0 - \sqrt{{\bf p}^2+\mu^2}\right)\left(p_0 + \sqrt{{\bf p}^2+\mu^2}\right)} \sim e^{-i\sqrt{{\bf p}^2+\mu^2}t} \sim e^{-iEt}e^{-(m/E)(\Gamma/2)t}
\end{equation}
where the energy $E$, identified by the oscillatory time dependence of the scattering amplitude, is
\begin{equation}
E= \frac{1}{\sqrt 2}\left({\bf p}^2+m^2-\Gamma^2/4+\sqrt{\left({\bf p}^2+m^2-\Gamma^2/4\right)^2+m^2\Gamma^2}\right)^{1/2}
\end{equation}
which reduces to $m$ when ${\bf p}=0$. As expected, the decay rate is reduced by a factor of $m/E$, which is the inverse of the usual time-dilation factor $\gamma = E/m$.

All of these results follow from the presence of a complex pole in the scattering amplitude, Eq.~(\ref{LOprop}). We introduced that pole via the propagator of an unstable particle, but there is also another possibility. Rather than introduce a field for the unstable particle, the unstable particle could arise as a composite of the other particles involved in the scattering amplitude (see Sec.~10.2 of Ref.~\cite{Weinberg}). This generally requires summing certain Feynman diagrams to all orders in perturbation theory, yielding a real (stable particle) or complex (unstable particle) pole. Such a particle is said to be dynamically generated. All of the results above for unstable particles continue to hold true.

One can go even further and dispense with Quantum Field Theory altogether, and simply regard a complex pole in the scattering amplitude as being associated with an unstable particle. This is the point of view of the $S$-matrix theory \cite{Chew} that was popular in the 1960s, before the renaissance of Quantum Field Theory.

Because scattering amplitudes are physical, we can draw some conclusions about the complex pole position associated with an unstable particle:
\begin{itemize}
\item The pole position is physical. Although the scattering amplitude has many other contributions to it in addition to the propagator, none of them effect the position of the pole.
\item Since the pole position is physical, it is invariant under field redefinitions.
\item Since the pole position is physical, it is gauge invariant (in a gauge theory).
\item The pole position is process independent. It doesn't matter what the initial or final states are, the pole position is the same.
\item The pole position is infrared safe. If the unstable particle carries electric or color charge, the pole position is unaffected by soft photons or gluons.
\end{itemize}
These are all highly desirable qualities of a definition of the mass and decay rate of an unstable particle.

\newpage

\noindent {\bf 3.\ Decay Rate}

\bigskip

Returning to Eq.~(\ref{pole}), and using Eq.~(\ref{mGamma}), we find
\begin{equation}
{\rm Im}\,\mu^2 = -m\Gamma = - {\rm Im}\, \Pi(\mu^2)
\end{equation}
or
\begin{equation}
\Gamma = \frac{1}{m}{\rm Im}\, \Pi(\mu^2)\;.
\label{width}
\end{equation}
This is an implicit formula for the decay rate, since $\mu = m - \frac{i}{2}\Gamma$. In many cases it can be solved by expanding in powers of $\Gamma/m$,
\begin{equation}
m\Gamma  = Z\,{\rm Im}\,\Pi(m^2)\left(1-\frac{1}{2}{\rm Im}\,\Pi(m^2){\rm Im}\,\Pi^{\prime\prime}(m^2)
-\frac{1}{4m^2}{\rm Im}\,\Pi(m^2){\rm Im}\,\Pi^\prime(m^2)+\cdots\right)
\label{mG}
\end{equation}
where the prime denotes differentiation and where $Z = [1+{\rm Re}\,\Pi^\prime(m^2)]^{-1}$ is the analogue of the field renormalization constant that would be necessary if the particle were an asymptotic state.

The leading term in Eq.~(\ref{mG}), $m\Gamma={\rm Im}\,\Pi(m^2)$, is the familiar leading-order expression for the decay rate, via the optical theorem (see Sec.~3.6 of Ref.~\cite{Weinberg}). That expression is usually derived by treating the decaying particle as an asymptotic state, which is clearly an approximation valid only for $\Gamma \ll m$. The exact expression, without any approximations, is given by Eq.~(\ref{width}).

The leading-order expression for the decay rate, $m\Gamma={\rm Im}\,\Pi(m^2)$, treats the decaying particle as if it is on shell at $p^2=m^2$. But an unstable particle is never truly on shell; on shell for an unstable particle corresponds to $p^2=\mu^2$, which is complex. There is nothing special about $p^2=m^2$ for an unstable particle.

\bigskip

\noindent {\bf 4. Width}

\bigskip

In the energy region near the pole at $p_0 = \mu$, the scattering amplitude of Eq.~(\ref{Epole}) can be approximated by neglecting the antiparticle pole at $p_0 = -\mu$,
\begin{equation}
{\cal M} \sim \frac{1}{E - \mu}
\end{equation}
where $E=p_0$.
The cross section in the resonance region is thus approximately
\begin{equation}
|{\cal M}|^2 \sim \frac{1}{(E-m)^2 + \Gamma^2/4}
\label{BW}
\end{equation}
which is the well-known Breit-Wigner formula. The resonance shape has a full width at half maximum of $\Gamma$, which is why the decay rate $\Gamma$ is also called the width. A similar formula may also be obtained from nonrelativistic quantum mechanics, such as in the original paper \cite{BW}. For this reason Eq.~(\ref{BW}) is sometimes referred to as a ``nonrelativistic Breit-Wigner,'' but we avoid this terminology since the equation is also valid at relativistic energies.

The physical process that Breit and Wigner addressed in their original 1936 paper \cite{BW}, entitled ``Capture of Slow Neutrons,'' is the absorption of a nonrelativitic neutron of kinetic energy $E$ by a nucleus, resulting in an excited nucleus of resonance energy $E_R$ that then decays to a lower-energy state by emission of a photon. They derived their resonance formula using nonrelativistic quantum mechanics,
\begin{equation}
|{\cal M}|^2 \sim \frac{1}{(E-E_R)^2 + \Gamma^2/4}\;.
\label{BW2}
\end{equation}
It is the similarity of Eq.~(\ref{BW}) to this equation that earns it the name Breit-Wigner.

Inserting Eq.~(\ref{mGamma}) into Eq.~(\ref{LOprop}) without making any approximations gives the amplitude
\begin{equation}
{\cal M} \sim \frac{1}{p^2 - \mu^2} = \frac{1}{p^2 - \left(m - \frac{i}{2}\Gamma\right)^2}
\label{propagator}
\end{equation}
and hence the cross section
\begin{equation}
\boxed{|{\cal M}|^2 \sim \frac{1}{(p^2 - m^2 + \Gamma^2/4)^2+m^2\Gamma^2}}\;.
\label{resonance}
\end{equation}
This resonance formula includes both the particle and antiparticle poles. For this reason it is manifestly Lorentz invariant, and it makes sense to refer to it as a ``relativistic Breit-Wigner,'' although those authors never wrote this formula down. It is common to see this formula in the literature in the approximation $\Gamma \ll m$, even when applied to particles for which this is not a good approximation, such as hadronic resonances or the $W$ and $Z$ bosons.

Another way to view the relativistic Breit-Wigner formula, Eq.~(\ref{resonance}), is to evaluate it in the rest frame of the particle and antiparticle,
\begin{equation}
|{\cal M}|^2 \sim \frac{1}{(E-m)^2 + \Gamma^2/4}\cdot\frac{1}{(E+m)^2 + \Gamma^2/4}
\end{equation}
which is most easily obtained directly from Eq.~(\ref{Epole}).
This makes it manifest that the relativistic Breit-Wigner formula is simply the product of a Breit-Wigner resonance for the particle and the antiparticle.

These Breit-Wigner formulae are based on an idealized world in which the only contribution to the amplitude is a pole. In the real, nonidealized world there are also nonresonant contributions to amplitudes.  The pole position continues to define the mass and decay rate via $\mu = m - \frac{i}{2}\Gamma$.

\bigskip

\noindent {\bf 5. Mass of the Roper Resonance}

\bigskip

What is the mass of the Roper resonance, the particle called $N(1440)$ by the Particle Data Group \cite{RPP2024}? You would be forgiven for thinking the answer is 1440 MeV. The physical mass, which we define by the real part of the pole in the complex energy plane, is approximately 1370 MeV according to the Particle Data Group.\footnote{We will have more to say about the mass of the Roper resonance at the very end of the paper.} So what, then, does 1440 MeV signify?

Rather than obtaining the mass from the pole in the propagator, Eq.~(\ref{pole}), let's define it as the zero of the real part of the denominator of the propagator,
\begin{equation}
M^2 - m_B^2 + {\rm Re}\,\Pi(M^2) = 0\;.
\label{mBW}
\end{equation}
For a stable particle these two definitions are identical, since $\Pi(\mu^2)$ is real for a stable particle. However, they differ for an unstable particle. This definition of mass is called the ``Breit-Wigner mass,'' although those authors had nothing to do with it. From the point of view of mathematics, Eq.~(\ref{mBW}) is quite strange, as there is nothing special about the point where the real part of the denominator of a complex function vanishes. In contrast, the pole position of a complex function is very special indeed. From the point of view of physics, Eq.~(\ref{mBW}) is a disaster. Recall that the pole position has the many desirable properties that we listed at the end of Sec.~2. The ``Breit-Wigner mass'' has none of them. Nevertheless, the ``Breit-Wigner mass'' caught on early in the history of hadronic resonances, and despite its many drawbacks it is still in use today. Almost all the hadronic resonances, including the Roper resonance, take the numerical part of their name from their ``Breit-Wigner mass,'' despite the fact that this mass is not well defined.\footnote{Another example is the $\Delta(1232)$, which has a physical mass of about 1210 MeV.} The $W$ and $Z$ boson masses are also defined by their ``Breit-Wigner masses,'' even though these masses are gauge dependent.\footnote{In the context of the $W$ and $Z$ bosons, the ``Breit-Wigner mass'' is also called the ``on-shell'' mass.}

The history of this strange situation is traced in Ref.~\cite{W2}, and we will not repeat it here. Fortunately the hadronic community has recognized the primacy of the pole position, and it is listed first, ahead of the ``Breit-Wigner mass,'' for most of the hadronic resonances recognized by the Particle Data Group \cite{RPP2024}. This is not the case for the $W$ and $Z$ bosons, however; you will not find the physical masses of the $W$ and $Z$ bosons listed. As we look forward to a future collider that can do ultraprecise $W$ and $Z$ boson physics, we will need to come to terms with this history.

\bigskip

\noindent {\bf 6. Poles and Branch Cuts}

\bigskip

In performing the Fourier transform of the amplitude in Sec.~2, we kept only the particle and antiparticle poles, and ignored any other singularities that might be present in the complex energy plane. We know that there are such singularities, in particular branch cuts, but they do not change the fact that a pole in the complex energy plane yields an exponential time dependence. In this section we discuss the branch cut associated with two-particle intermediate states in the propagator.

Let's reconsider the equation for the pole position $\mu^2$, Eq.~(\ref{pole}). Since the self energy is real analytic, $\Pi^{*}(p^2)=\Pi(p^{2*})$, taking the complex conjugate of Eq.~(\ref{pole}) yields
\begin{equation}
\mu^{2*}-m_B^2+\Pi(\mu^{2*})=0
\end{equation}
which shows that there is also a pole at $\mu^{2*}$, which we will refer to as the complex-conjugate pole.
Now consider the self energy on the real axis above the threshold for a two-particle intermediate state (or, more generally, a multiparticle intermediate state). Unitarity demands that the imaginary part of $\Pi(p^2)$ is proportional to the squared amplitude for the creation of the two-particle state, times the two-particle phase space (see Sec.~3.6 of Ref.~\cite{Weinberg}). The phase space for two particles is proportional to $q^{2L+1}$, where $q$ is the momentum of each of the two particles in the center-of-momentum frame and $L$ is their angular momentum. The explicit expression for $q$ is
\begin{equation}
q=\frac{1}{2\sqrt{p^2}}\sqrt{[p^2-(m_1+m_2)^2][p^2-(m_1-m_2)^2]}
\label{q}
\end{equation}
where $m_1,m_2$ are the masses of the intermediate particles. Note the square root in the expression for $q$; this means that $\Pi(p^2)$ is a double-valued function, with a branch point at the two-particle threshold, $p^2=(m_1+m_2)^2$. There is also a branch point at $p^2=(m_1-m_2)^2$, dubbed a ``pseudothreshold.''

The analytic structure of the momentum-space propagator in the complex $p^2$ plane is shown in Fig.~3 (see also Sec.~50 of Ref.~\cite{RPP2024}).  There is a branch point at the threshold, with a branch cut extending to infinity along the positive real $p^2$ axis.  There is a new branch point and branch cut for each new threshold, but we show only the lowest threshold for clarity of presentation. The branch cut continuously connects the two branches, or Riemann sheets, of the propagator. The first sheet is defined by ${\rm Im}\,q>0$, and the second sheet by ${\rm Im}\,q<0$. The pseudothreshold lies on the second sheet, with a branch cut that extends to negative infinity along the real $p^2$ axis (not shown in the figure). The physical axis lies on the first sheet, just above the branch cut. The pole in the propagator lies on the second sheet at $p^2 = \mu^2$, which is accessed by passing downward through the branch cut from the physical axis.  The complex-conjugate pole at $p^2 = \mu^{2*}$ is also on the second sheet, but it is much further from the physical axis than the pole at $p^2=\mu^2$, since one needs to circle the branch point to get from the complex-conjugate pole to the physical axis.

The analytic structure of the momentum-space propagator in the complex energy ($p_0$) plane, with ${\bf p} = 0$, is shown in Fig.~4. There is a branch point at the threshold on the positive real axis as well as on the negative real axis.  There is a particle pole on the second sheet below the branch cut on the positive real axis, and an antiparticle pole on the second sheet above the branch cut on the negative real axis.  As in the $p^2$ plane, there are also poles at the complex-conjugate positions, far from the physical axis.  Also shown is the contour of integration for the Fourier transform of the propagator.

The Fourier transform can be performed by closing the contour on the first sheet, as shown in Fig.~5. Since the poles lie on the second sheet, they are not shown in the figure, and are not enclosed by the contour.  Let the momentum-space propagator of Eq.~(\ref{bare2}) be denoted by
\begin{equation}
G(p_0) = \frac{i}{p_0^2 - \mu^2 + \Pi(p_0^2) - \Pi(\mu^2)+i\epsilon} \;.
\end{equation}
Then, for $t>0$, the contour integration gives
\begin{equation}
\int_{-\infty}^\infty \frac{dp_0}{2\pi}\; e^{-ip_0t}G(p_0) = \int_{thr}^\infty \frac{dp_0}{2\pi}\; e^{-ip_0t}\;{\rm Disc}\;G(p_0)
\label{Disc}
\end{equation}
where ${\rm Disc}\;G(p_0)$ is the discontinuity of $G(p_0)$ across the branch cut running from threshold to infinity. Let's compare this with the Fourier transform of the K\"allen-Lehmann representation of the propagator (see Sec.~6.1 of Ref.~\cite{Brown} or Sec.~10.7 of Ref.~\cite{Weinberg}),
\begin{equation}
\int_{-\infty}^\infty \frac{dp_0}{2\pi}\; e^{-ip_0t}G(p_0) = \int_{-\infty}^\infty \frac{dp_0}{2\pi}\; e^{-ip_0t} \int_{thr^2}^\infty \frac{dM^2}{2\pi}\rho(M^2)\frac{i}{p_0^2-M^2+i\epsilon}
\end{equation}
where $\rho(M^2)$ is the spectral function. For $t>0$ we close the $p_0$ contour in the lower half plane and pick up the residue of the pole at $p_0 = M-i\epsilon$, yielding
\begin{equation}
\int_{-\infty}^\infty \frac{dp_0}{2\pi}\; e^{-ip_0t}G(p_0) = \int_{thr}^\infty \frac{dM}{2\pi}\; e^{-iMt}\rho(M^2)\;.
\end{equation}
Comparing with Eq.~(\ref{Disc}) we conclude
\begin{equation}
\rho(p_0^2) = {\rm Disc}\;G(p_0)\;.
\end{equation}
Note that the spectral function does not have a delta function contribution from the unstable particle, as it would from a stable particle, because the unstable particle is not part of the spectrum of asymptotic states.

The Fourier transform performed by closing the contour on the first sheet yields an integral over the discontinuity in the propagator, Eq.~(\ref{Disc}), but it does not explicitly yield exponential time dependence, since the contour does not enclose any poles. Nevertheless, the exponential time dependence must arise from the integration. The pole on the second sheet causes the imaginary part of the propagator on the first sheet to have a peak above the branch cut, and the complex-conjugate pole on the second sheet causes the imaginary part of the propagator on the first sheet to have valley below the branch cut (see Sec.~50 of Ref.~\cite{RPP2024}), so there is a large discontinuity across the branch cut that must yield exponential time dependence.

To see how this works, let's approximate the propagator just above the branch cut by the pole, and approximate the propagator just below the branch cut by the complex-conjugate pole. Thus
\begin{equation}
{\rm Disc}\;G(p_0) \sim
\frac{i}{p_0-\mu}-\frac{i}{p_0-\mu^*}
=\frac{\Gamma}{(p_0-m)^2+\Gamma^2/4}\;.
\end{equation}
We now perform the Fourier transform on the right-hand side of Eq.~(\ref{Disc}), extending the lower limit of integration to negative infinity since the integration region to the left of the threshold is far from the poles. This yields the exponential time dependence we anticipated,
\begin{equation}
\int_{-\infty}^\infty \frac{dp_0}{2\pi}\; e^{-ip_0t}\;\frac{\Gamma}{(p_0-m)^2+\Gamma^2/4}
=e^{-imt}e^{-(\Gamma/2)|t|}
\end{equation}
where we have used the well-known formula for the Fourier transform of a Lorentzian function. The reader will notice that the integrand contains the formula for the Breit-Wigner cross section, Eq.~(\ref{BW}), but recall that we are evaluating the Fourier transform of the amplitude, not the cross section.

To make the exponential time dependence from the pole explicit without any approximations, it is necessary to close the contour on the second sheet, as shown in Fig.~6, rather than on the first sheet as was done above. One finds
\begin{equation}
\int_{-\infty}^\infty \frac{dp_0}{2\pi}\; e^{-ip_0t}G(p_0)
= Re^{-i\mu t} + \int_{-\infty}^{thr} \frac{dp_0}{2\pi}\; e^{-ip_0t}\;{\rm Disc}\;G(p_0)
\label{secondsheet}\end{equation}
where $R = (2\mu[1+\Pi^\prime(\mu^2)])^{-1}$ is the residue of the pole. This alternative evaluation of the Fourier transform makes the exponential time dependence associated with the unstable particle pole explicit. As the contour on the large semicircle passes the negative imaginary axis in Fig.~6 it crosses the pseudothreshold branch cut (not shown in the figure)\footnote{There is a branch cut from the pseudothreshold at $p_0 = |m_1-m_2|$ on the second sheet that extends to the origin and then up the positive imaginary axis, and a branch cut from the pseudothreshold at $p_0 = -|m_1-m_2|$ on the second sheet that extends to the origin and then down the negative imaginary axis.} onto a third sheet, where there are no poles. Although the contour appears to enclose a pole in the lower left quadrant, that pole is on the second sheet, while the contour is on the third sheet in that quadrant. The discontinuity in Eq.~(\ref{secondsheet}) is between the first and third sheets from negative infinity up to the origin, and between the first and second sheets from the origin up to the threshold.

\newpage

\noindent {\bf 7. Pole Trajectories}

\bigskip

Let's follow a pole as it moves in the complex $p^2$ plane from being a stable particle to an unstable particle, by varying some parameter \cite{BhW,HPR}. We study a very simple theory of a scalar particle $S$ coupled to a pair of scalar particles $A$, such that the threshold is at $p^2=4m_A^2$. We also couple $S$ to another pair of scalar particles $B$, such that there is a second, higher threshold at $p^2=4m_B^2$, so we can study the movement of the pole near a threshold. The parameter we vary is the ``Breit-Wigner mass'' of $S$, with the caveat that this parameter is not invariant under field redefinitions and is unphysical. We will simply refer to it as the ``mass parameter,'' to emphasize that it is not the physical mass of $S$.

The pole trajectories in the complex $p^2$ plane are shown in Fig.~7, taken from Ref.~\cite{BhW}. Both the poles and their complex conjugates are shown. We are using the plural word ``poles'' because there is more than one, as we will discuss. The branch points at $4m_A^2$ and $4m_B^2$ are shown (we will refer to them as branch points $A$ and $B$), and we have rotated the branch cuts downward by $90^{\circ}$ in order to expose the second sheets in the lower half plane with respect to these branch points.

Let's begin with the pole labeled by the mass parameter 350 (in arbitrary units). In this case the mass of $S$ actually is 350, and it is stable against decay into pairs of $A$ ($m_A = 200$) or $B$ ($m_B = 250$) particles, so the pole is on the real axis on the first sheet. We increase the mass parameter to 375 (pole shown but not labeled in the figure) and then to 400, where it sits on top of branch point $A$, still a stable particle. Note, however, that there are also a pole and its complex conjugate labeled 400 on the left side of the figure, lying on the second sheet. These are far from the physical axis, which begins at branch point $A$ and lies on the first sheet just above the positive real axis.

As we increase the mass parameter further, the pole drops through the branch point onto the second sheet and moves {\it backwards}. This surprising behavior only happens for $L=0$. The result is a pole on the real axis below the branch point and on the second sheet. This corresponds to a stable particle that is not an asymptotic state. The $S$ particle cannot be an external line in a Feynman diagram, but it can be an internal line coupled to pairs of $A$ or $B$ particles. It is impossible for an internal $S$ line to be on shell, since its mass is less than $4m_A^2$. Such a particle is called a virtual state.\footnote{Although it is called a virtual state, it is not an asymptotic state. The same is true of an unstable particle, which is often referred to as an unstable state.} Remarkably, this seemingly exotic state exists in the two-nucleon system. There is the well-known $J^P=1^+$, $I=0$ bound state, namely the deuteron, but there is also a $J^P=0^+$, $I=1$ virtual bound state. Both of these are examples of dynamically-generated states.

If we increase the mass parameter further, the pole moves off the real axis into the complex plane on the second sheet. To be more precise, as the mass parameter is increased the pole and its complex conjugate labeled 400 on the left side of the figure move towards each other and meet on the real axis, and then one goes left and one goes right. The one going right collides with the backwards-moving pole, and they then move off the real axis as a pole and its complex conjugate. This pole correspond to an unstable particle, even though the physical mass of the particle (the real part of the pole in the complex energy plane) is less than $2m_A$. This apparent paradox is resolved if we recall that an unstable particle should never be thought of as being on shell at some real value of $p^2$. An unstable particle, like a virtual state, is always off shell and never an external particle. As we mentioned at the end of Sec.~3, there is nothing special about $p^2=m^2$ for an unstable particle; an unstable particle does not have to satisfy $m^2>4m_A^2$ (for $L=0$).

As we increase the mass parameter further to 425 (pole shown but not labeled), the unstable-particle pole crosses the rotated branch cut, and can be seen moving on the exposed second sheet to 450 and 475 (pole shown but not labeled). When we increase the mass parameter to 500, the pole crosses the rotated branch cut associated with branch point $B$. The pole is still on the second sheet of branch point $A$, and it is still on the first sheet of branch point $B$. Meanwhile, a pole and its complex conjugate, also labeled 500, come in from the left on an outer trajectory on the second sheet of $A$ and the second sheet of $B$. This pole crosses the rotated branch cut associated with branch point $B$ at mass parameter 525 (pole shown but not labeled), and continues on to 550. This pole at 550 is close to the physical axis, as it is on the second sheet with respect to both branch points, so it is directly below the physical axis. The pole labeled 550 on the inner trajectory, although seemingly close to the physical axis, is far away from it, since it is on the first sheet with respect to branch point $B$, so one must circle around this branch point to arrive at the physical axis.

Consider the poles in the range 500 to 525 that lie below branch point $B$. The poles on the inner trajectory and the outer trajectory are both fairly close to the physical axis. They will both make their presence known in the scattering amplitude on the physical axis. Are these two distinct particles, or one? I think the right attitude is that it is one particle with two poles. After all, these two poles are not independent of each other, as they would be if they were from two distinct particles. But what then is the mass and width of this particle? We must conclude that it has two masses and widths. While this seems like an exotic situation, it is actually not uncommon to find hadronic resonances near a threshold. We need look no further than the Roper resonance for an example; it lies near the $\pi\Delta(1232)$ threshold and has two poles associated with it, one on the first sheet and one on the second sheet with respect to the $\pi\Delta(1232)$ branch point, a fact uncovered by Dave and collaborators twenty years after the discovery of the particle \cite{AFR}. Thus the Particle Data Group should list two pole positions for the Roper resonance. Particles near a threshold, including the Roper resonance \cite{DH2KM,RDH4KMN,KHKS}, are candidates for dynamically-generated states.

\bigskip

\noindent {\bf Acknowledgements}

\bigskip

I am grateful for correspondence with Tanmoy Bhattacharya, Christoph Hanhart, and Ulf Mei{\ss}ner.

\begin{figure}
\centering
\includegraphics[width=\textwidth]{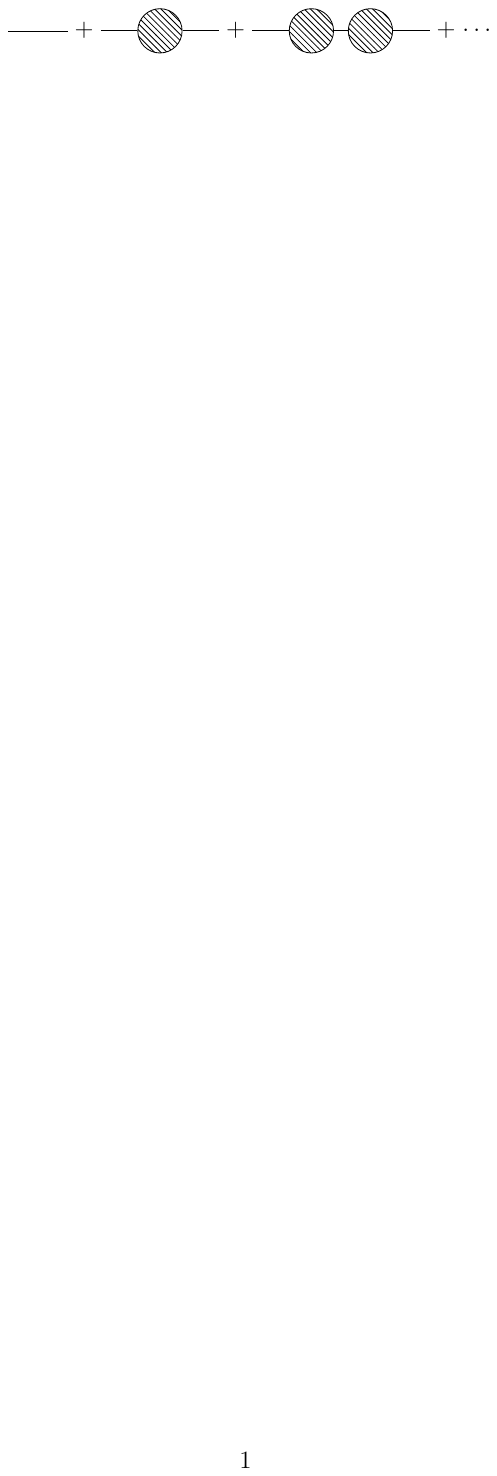}
\caption{Self-energy corrections to the propagator.}
\end{figure}

\begin{figure}
\centering
\includegraphics[width=\textwidth]{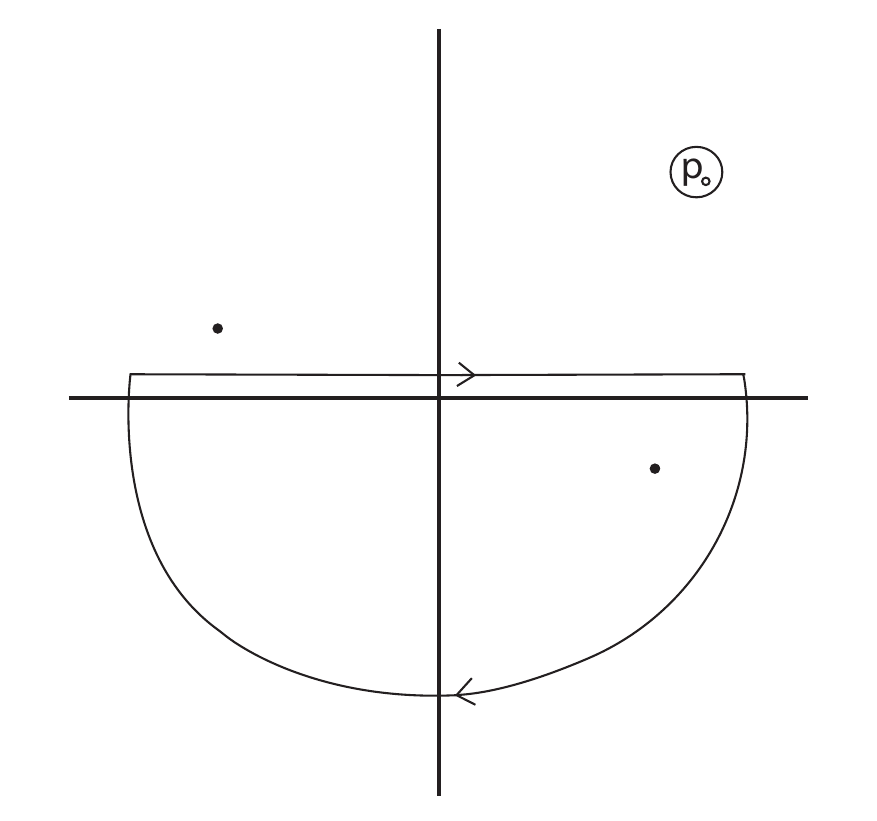}
\caption{Evaluating the Fourier transform of the propagator by closing the contour in the lower-half energy plane (for $t>0)$.}
\end{figure}

\begin{figure}
\centering
\includegraphics[width=\textwidth]{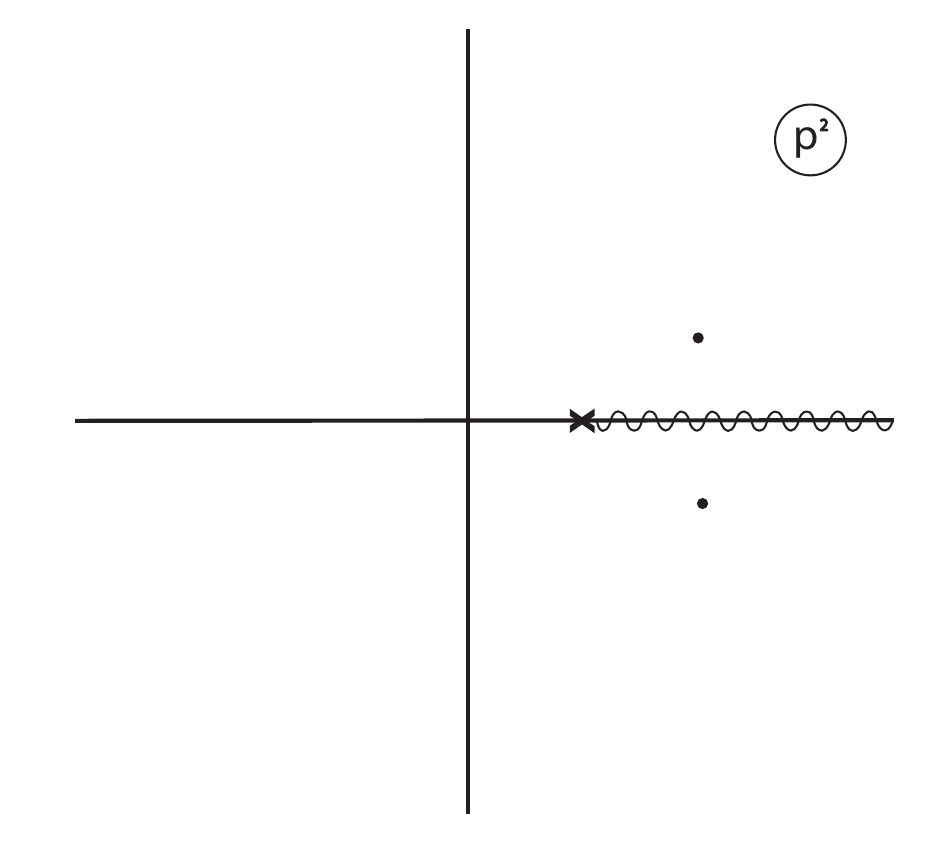}
\caption{Analytic structure of the propagator in the complex $p^2$ plane. The poles lie on the second sheet.}
\end{figure}

\begin{figure}
\centering
\includegraphics[width=\textwidth]{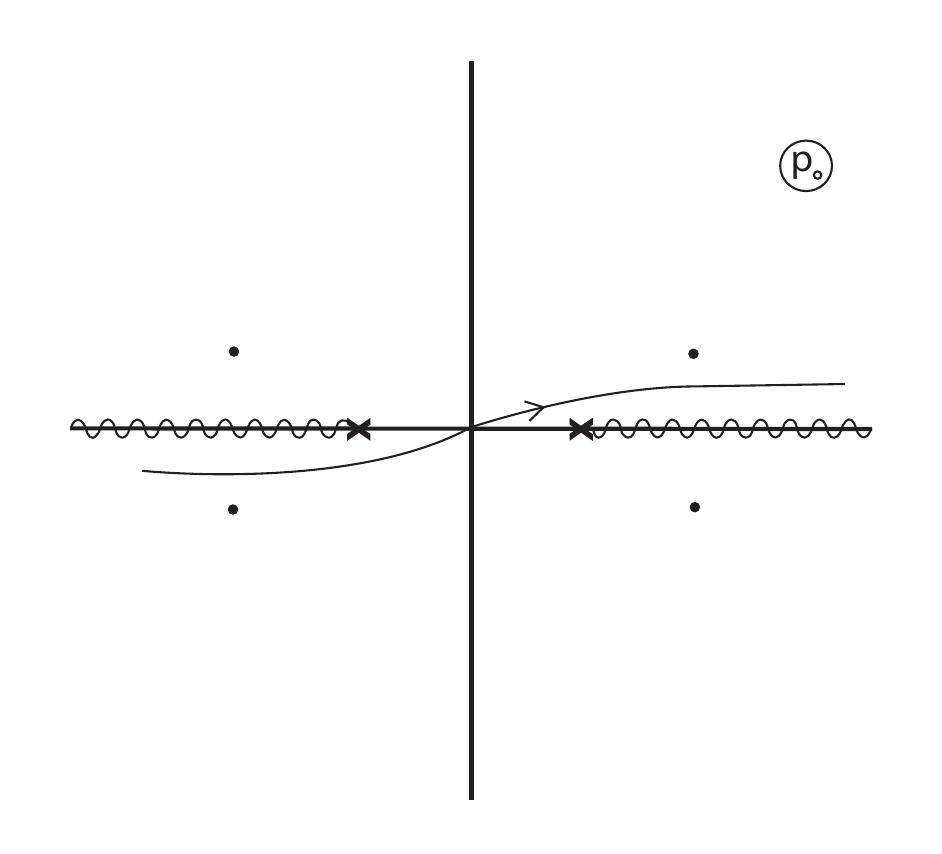}
\caption{Analytic structure of the propagator in the complex energy plane. The poles lie on the second sheet. The integration contour for the Fourier transform is also shown.}
\end{figure}

\begin{figure}
\centering
\includegraphics[width=\textwidth]{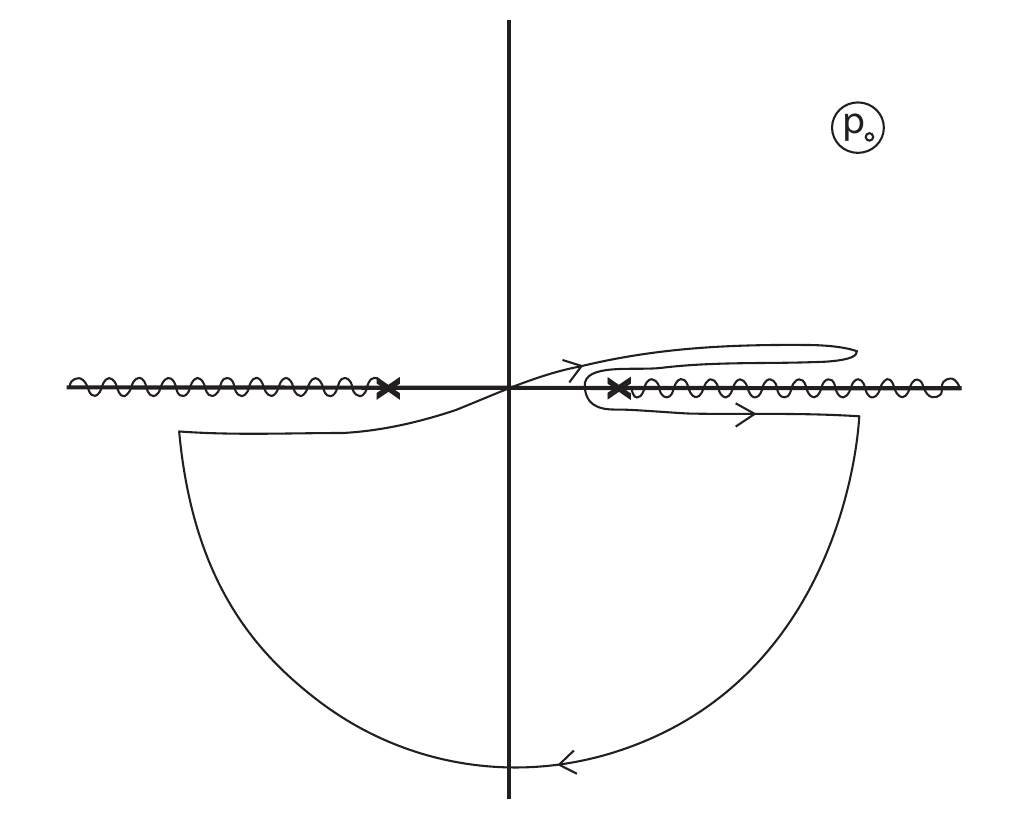}
\caption{Evaluating the Fourier transform of the propagator by closing the contour in the lower-half energy plane (for $t>0)$ on the first sheet.}
\end{figure}

\begin{figure}
\centering
\includegraphics[width=\textwidth]{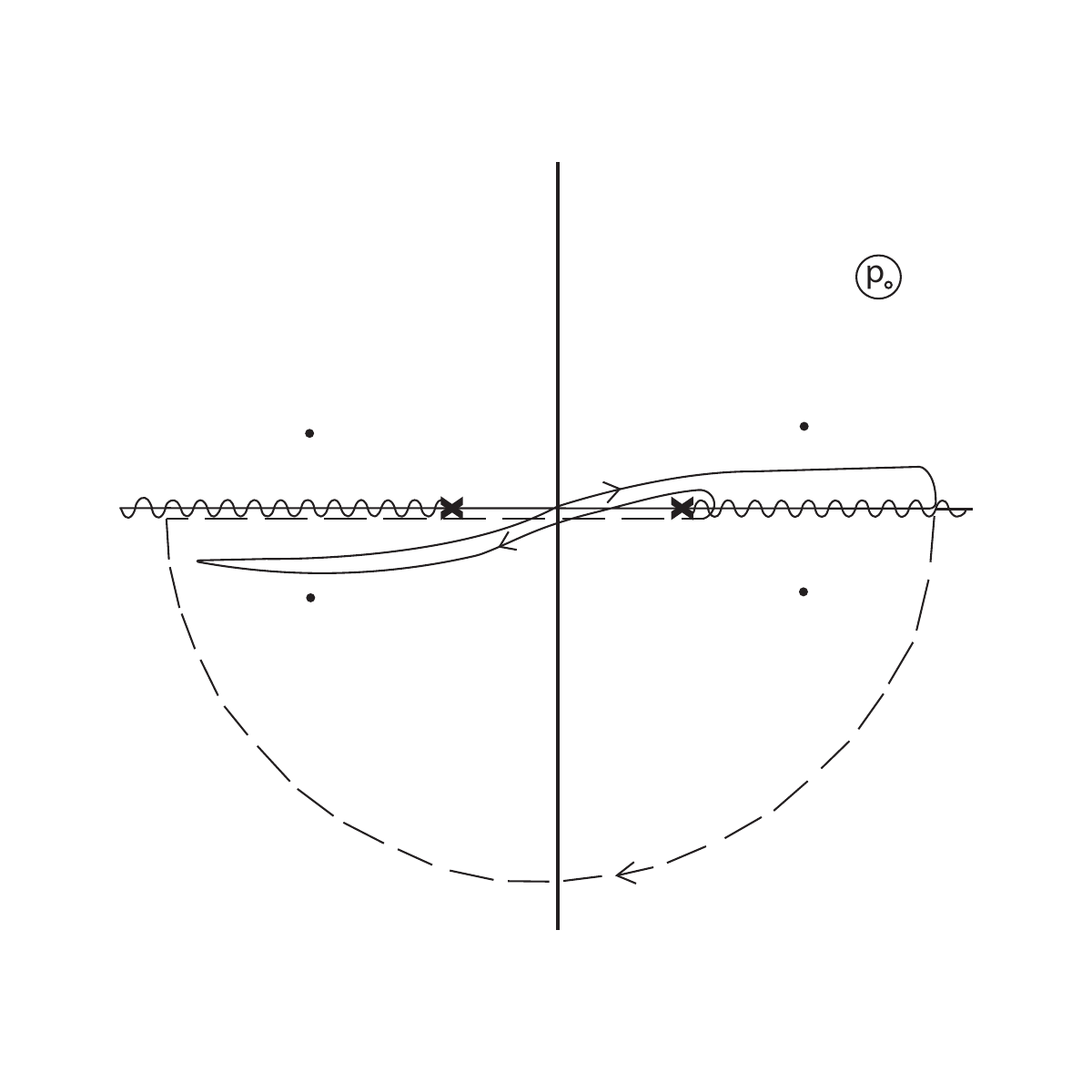}
\caption{Evaluating the Fourier transform of the propagator by closing the contour in the lower-half energy plane (for $t>0)$ on the second sheet.}
\end{figure}

\begin{figure}
\centering
\includegraphics[width=\textwidth]{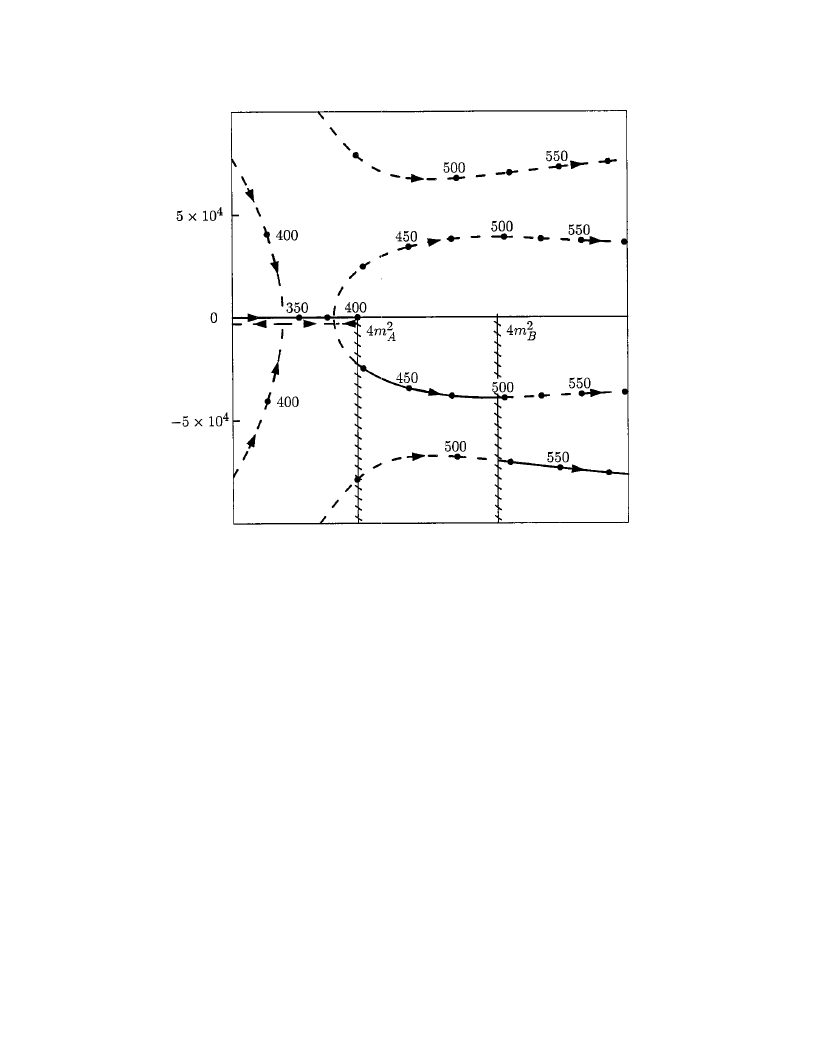}
\caption{Poles in the complex $p^2$ plane as a function of the mass parameter, indicated by the dots, for $m_A = 200$ and $m_B = 250$ (arbitrary units). The arrows indicated the direction of motion of the poles as the mass parameter is increased. The solid curves lie on exposed sheets; all other curves are covered by one or two sheets.}
\end{figure}


\newpage

\begin{thebibliography}{100}

\bibitem{W1} S.~Willenbrock, Eur.\ Phys.\ J.\ Plus {\bf 139}, 523 (2024).

\bibitem{W2} S.~Willenbrock, Eur.\ Phys.\ J.\ Plus {\bf 140}, 502 (2025).

\bibitem{Brown} L.~Brown, {\sl Quantum Field Theory} (Cambridge University Press, 1992).

\bibitem{Roper} L.~David Roper, Phys.\ Rev.\ Lett.\ {\bf 12}, 340 (1964); L.~David Roper, R.~Wright, and B.~Feld, Phys.\ Rev.\ {\bf 138}, B190 (1965).

\bibitem{RPP2024} S.~Navas {\it et.\ al.} (Particle Data Group), Phys.\ Rev.\ D {\bf 110}, 030001 (2024).

\bibitem{Weinberg} S.~Weinberg, {\sl The Quantum Theory of Fields, Vol.~I} (Cambridge University Press, 1995).

\bibitem{Chew} G.~Chew, {\sl The Analytic S Matrix} (W.~A.~Benjamin, 1966).

\bibitem{BW} G.~Breit and E.~P.~Wigner, Phys.\ Rev.\ {\bf 49}, 519 (1936).

\bibitem{BhW} T.~Bhattacharya and S.~Willenbrock, Phys.\ Rev.\ D {\bf 47}, 4022 (1993).

\bibitem{HPR} C.~Hanhart, J.~Pelaez, and G.~Rios, Phys.\ Lett.\ B {\bf 739}, 375 (2014).

\bibitem{AFR} R.~Arndt, J.~Ford, and L.~David Roper, Phys.\ Rev.\ D {\bf 32}, 1085 (1985).

\bibitem{DH2KM} M.~D\"oring, C.~Hanhart, F.~Huang, S.~Krewald, and U.-G.~Mei{\ss}ner, Nucl.\ Phys.\ A {\bf 829}, 170 (2009).

\bibitem{RDH4KMN} D.~R\"onchen, M.~D\"oring, F.~Huang, H.~Haberzettl, J.~Haidenbauer, C.~Hanhart, S.~Krewald, U.-G.~Mei{\ss}ner, and K.~Nakayama, Eur.\ Phys.\ J.\ A {\bf 49}, 44 (2013).

\bibitem{KHKS} O.~Krehl, C.~Hanhart, S.~Krewald, and J.~Speth, Phys.\ Rev.\ C {\bf 62}, 025207 (2000).

\end{thebibliography}
\end{document}